\documentclass[aps,floatfix,twocolumn,a4paper,showpacs, nofootinbib,superscriptaddress,10pt]{revtex4}
\usepackage{graphicx,float}\usepackage{graphicx,float}
\usepackage[all]{xy}
\usepackage{amsmath,upgreek}
\usepackage{amssymb}
\usepackage{color}
\usepackage{epsfig}		
\usepackage{graphicx,epstopdf}
\usepackage{subfigure}
\usepackage{pdfpages}
\usepackage[colorlinks,hyperindex]{hyperref}
 \newcommand{\blt}{\textcolor{black}}
\definecolor{green1}{RGB}{0,128,0} 
\hypersetup{backref=true,pagebackref=true}
\hypersetup{%
  colorlinks = true,
  linkcolor  = blue,
  citecolor = cyan,
}
\usepackage{bookmark,textgreek}
\usepackage{hyperref,color,xcolor}
\hypersetup{hidelinks,hyperindex=true,colorlinks=true,breaklinks=true,urlcolor= blue}
\hypersetup{%
  colorlinks = true,
  linkcolor  = blue
}
\usepackage{graphicx,float,tikz}
\usepackage[all]{xy}
\newcommand\ringring[1]{%
  {
   \mathop{\kern0pt #1}\limits^{
     \vbox to-1.85ex{
       \kern-2ex 
       \hbox to 0pt{\hss\normalfont\kern.1em \r{}\kern-.45em \r{}\hss}%
       \vss 
     }
   }
  }
}\newcommand\orcidroldao{{\href{https://orcid.org/0000-0003-3978-532X}{\orcidicon}}}
\newcommand{\orcidicon}{%
	\begin{tikzpicture}
	\draw[lime, fill=lime] (0,0)
		circle [radius=0.16]
		node[white] {{\fontfamily{qag}\selectfont \tiny ID}};
	\draw[white, fill=white] (-0.0625,0.095)
		circle [radius=0.007];
	\end{tikzpicture}	\hspace{-2mm}
}

\newcommand{\bes}{\begin{subequations}}
\newcommand{\ees}{\end{subequations}}
\def\beq{\begin{eqnarray}}
 
\def\eeq{\end{eqnarray}}
\def\be{\begin{equation}}
\def\ee{\end{equation}}

\begin{document}

\title{Deploying heavier $\eta$ meson states: configurational entropy hybridizing AdS/QCD}
\author{R. da Rocha\orcidroldao\!\!}
\email{roldao.rocha@ufabc.edu.br}
\affiliation{Federal University of ABC, Center of Mathematics, Santo Andr\'e, 09580-210, Brazil}

\begin{abstract}
The meson family of $\eta$ pseudoscalars is studied in the context of the AdS/QCD correspondence and the differential configurational entropy (DCE). For it, two forms of  configurational-entropic Regge-like trajectories are engendered,  relating the $\eta$ mesonic states excitation number to both their experimental mass spectrum in the Particle Data Group (PDG) and the DCE as well. 
Hence, the mass spectrum of $\eta$ pseudoscalar mesonic states, beyond the already detected states 
$\eta(550)$, 
	 $\eta(1295)$, $\eta(1405)$, $\eta(1475)$, $\eta(1760)$, $\eta(2225)$, and $\eta(2320)$, is derived for any excitation number. The three first ulterior members of this family are  then  analyzed and also compared to existing candidates in PDG.

 \end{abstract}
\pacs{89.70.Cf, 11.25.Tq, 14.40.Aq }
\maketitle

\section{Introduction}

Shannon's information entropy paradigm resides in converting information into a coded form in random systems. Compressed messages do present the same amount of information as the original ones. Nevertheless, they are less redundant \cite{Shannon:1948zz}. In his seminal work, Shannon replaced information, from an unestablished  concept, into a precise theory that has underlain the digital revolution. It has inspired the development of all modern data-compression algorithms and error-correcting codes as well. 
Information entropy is the core  of formulating the configurational entropy (CE) and consists of the part of the entropy of a system that is related to representative correlations among its finite number of integrants. The CE of any  system evaluates the rate at which information is condensed into wave modes. 
For the continuous limit, the differential configurational entropy (DCE) takes place and  can be applied to a considerable variety of  problems when one introduces some distribution portraying the system that is related to physical quantities under scrutiny \cite{Gleiser:2018kbq,Gleiser:2011di,Bernardini:2016hvx}.

The DCE merges informational and dynamical constituents of physical systems \cite{Gleiser:2012tu}. 
The DCE logarithmically evaluates the number of bits needed to designate the organization of the studied system. 
Employing the DCE to approach a system requires a  scalar field that is spatially localized. One  typically uses  the energy density, as the $T_{00}$ component of the energy-momentum tensor. Other choices, like scattering amplitudes, are also suitable, according to the system to be approached \cite{Karapetyan:2018oye,Karapetyan:2018yhm}. 
 The DCE was also employed to distinguish dynamical systems among regular, entirely random, and also chaotic evolution in QCD, providing observables and techniques
to study  physical phenomena \cite{Ma:2018wtw}.
 The DCE setup additionally refined the comprehension of phenomenological AdS/QCD, playing the 
color glass condensate an important role on studying meson states  \cite{Karapetyan:2016fai,Karapetyan:2017edu,Karapetyan:2019fst}. With the DCE tools, heavy-ion interactions were else investigated  \cite{Ma:2018wtw}. 
The DCE has been systematically shown to be a very important criterion to examine some of the most essential features of AdS/QCD. 

The DCE estimated new features of light-flavor mesonic  states of several families in AdS/QCD \cite{Bernardini:2016hvx,Bernardini:2018uuy,Ferreira:2019inu}, corroborated by several experiments reported in Particle Data Group (PDG) \cite{pdg}. 
Besides, charmonia and bottomonia, tensor mesons, scalar glueballs, odderons, pomerons, and baryons, amongst other particle states, were scrutinized in the DCE and AdS/QCD contexts, also incorporating finite temperature \cite{Ferreira:2019nkz,Colangelo:2018mrt,Ferreira:2020iry,Braga:2018fyc,Braga:2020myi,Braga:2020hhs}. A variety of leading applications of the DCE have been also investigated in Refs.  \cite{Correa:2016pgr,Cruz:2019kwh,Lee:2019tod,Bazeia:2018uyg,Alves:2014ksa,Alves:2017ljt,Gleiser:2018jpd}.
Besides, the DCE is decisive for investigating the gravity point of view in AdS/QCD duality. Black branes  were  studied in Ref.  \cite{Braga:2019jqg} in the context of the DCE, whereas black holes and thermal AdS backgrounds yielded a  temperature-dependent DCE in Ref. \cite{Braga:2020opg}.  Also, Bose-Einstein condensates in AdS/CFT, modeling self-interacting gravitons, were explored under the DCE apparatus \cite{Casadio:2016aum}.

The AdS/QCD framework is constituted by an AdS  bulk wherein gravity, which is weakly coupled, emulates the dual setup to the four-dimensional (conformal) field theory (CFT). In the duality dictionary, physical fields in the AdS bulk are dual objects to boundary operators of (strongly coupled) QCD.
The bulk fifth dimension is nothing more than the energy scale of  QCD. Confinement can be, hence, achieved by either by a Heaviside cut-off in the bulk, the hard wall,  or by a dilatonic field, that accomplishes a smooth cut-off along the AdS bulk -- the soft wall model \cite{Csaki,Karch:2006pv}. The soft wall model asserts that Regge trajectories naturally emerge from confinement when one employs a quadratic dilaton in the action that governs the bulk fields.  Although the soft wall model succeeds in some aspects, it does not address other relevant aspects of QCD. Those aspects that are not encompassed by the soft wall include the derivation of mesonic decay constants, the chiral symmetry breaking (CSB) and the quarkonium mass spectrum, among others  \cite{Contreras:2018hbi}. A decisive answer to these issues yielded a ultraviolet (UV) cut-off to be considered  \cite{Braga:2015jca}, permitting an energy scale $z_0$ sitting along the bulk. Although pseudoscalar mesons were scrutinized within a non-perturbative setup, in AdS/CFT  \cite{Katz:2007tf}, here a modified AdS/QCD model will be employed, where $\eta$ pseudoscalar mesons are dual objects to free bulk fields, having an appropriate anomalous dimension that gauge the bulk field mass \cite{Braga:2016wkm}.

In this work, the DCE for the $\eta$ pseudoscalar mesons   family will be computed, as a function of both the $\eta$ meson family excitation number and mass spectrum. The interpolation curves to these data engender two forms of DCE Regge-like trajectories. Hence, the mass spectrum of $\eta$ meson states, with higher excitation numbers, can be then obtained, consisting of the next generation of $\eta$ mesons to be experimentally detected. 
This article is displayed as follows: Sec. \ref{sec1}  devotes  to 
review the AdS/QCD duality. From the CSB and the introduction of the anomalous dimension, the mass spectrum of $\eta$ pseudoscalar mesons  family is derived, with the aid of bulk-to-boundary propagators and correlators. In Sec. \ref{sec2}, the fundamentals of the DCE are introduced and discussed, being the DCE computed for the $\eta$ pseudoscalar meson    family. The DCE is then expressed as a function of the $\eta$ mesons  $n$ excitation number and their experimental mass spectrum as well. Therefore, 
two forms of DCE Regge-like trajectories of $\eta$ mesons  consist of the resulting interpolation curves for these two functions, predicting the mass spectrum of heavier $\eta$ mesonic states, which do correspond to higher $n$  mesonic $\eta$ states.  Sec. \ref{iv} encompasses an overall analysis of the results and conclusions.
\vspace*{-.5cm}

\section{AdS/QCD essential framework}
\label{sec1}
The starting point is the model introduced in \cite{Braga:2015jca}. One defines the AdS Poincar\'e patch with UV cut-off as 
\begin{eqnarray}
ds^2&=&g_{AB}\,dx^A\,dx^B\\ \notag
&=&\frac{R^2}{z^2}\left(dz^2+\upeta_{\mu \nu}\,dx^\mu\,dx^\nu\right)\Uptheta\left(z-z_0\right),	
\end{eqnarray}
for $\Uptheta\left(x\right)$ being the Heaviside step function, $R$ is the AdS radius, and $z_0$ is the UV cut-off defined by the geometrical locus of a D-brane. Choosing the AdS boundary yields the conformal invariance to be naturally broken by the energy scale $z_0$ \cite{Braga:2015jca}. 
This model has been consistently employed in studying,  in particular, meson families in AdS/QCD \cite{Colangelo:2008us}. Making the conformal boundary $z_0 \to 0$ limit, the mass spectrum of mesons satisfies the standard Regge trajectory for the dilaton $\Upphi(z) = k^2z^2$. 

Pseudoscalar mesons can be described by a function $P(x^\mu,z)$, and the respective action reads 
\beq\label{123}
 S_{P}=-\frac{1}{2g_P^2}\int d^5x\,\sqrt{-g}{\scalebox{1}{$L$}},
 \eeq
 where the Lagrangian is given by 
\beq\label{acao}
{\scalebox{1}{$L$}}=e^{-\Upphi(z)}\left[g^{AB}\,\partial_AP\partial_BP+\mathcal{M}_5^2P^2\right],
\eeq
The constant $g_P$ fixes the units of the action with respect to the number of colors, $N_c$, of the CFT, whereas $\mathcal{M}_5^2$ stands for the bulk mass of pseudoscalar mesons. The action (\ref{acao}) yields  the bulk fields equations of motion. According to the gauge/gravity setup, scalar bulk fields solutions scale as $z^{\Updelta-4}$ in the UV regime, being dual to an  $\mathcal{O}$  operator in the boundary CFT. Its associated 2-point correlator reads \cite{Witten:1998qj} 
\begin{equation}
\left\langle\mathcal{O}(x)\,\mathcal{O}(\mathbf{0})\right\rangle \propto \left|x\right|^{-2\Updelta},
\end{equation}
where $\Updelta$ denotes the scaling dimension of  $\mathcal{O}$. The latter $\Updelta$ term is also related to the operators that create hadrons. Mesons can be thus implemented when one employs $q\bar{q}$ operators, corresponding to $\Updelta=3$. 
As bulk mass terms in the action affect the UV limit of the solutions, therefore $\Updelta$, the term $\mathcal{M}_5^2$, and the spin $s$ are coupled to as  \cite{Witten:1998qj} 
\begin{equation}\label{msd}
    \mathcal{M}_5^2\,R^2=\Updelta\,\left(\Updelta-4\right)-s\left(s-4\right).
\end{equation}
Scalar mesons correspond to $\mathcal{M}_5^2\,R^2=-3$, whereas vector mesons satisfy $\mathcal{M}_5^2\,R^2=0$. Eq. (\ref{msd}) holds for the $s$-wave, $\ell=0$, approach of mesons. For $\ell\neq0$ states, a twist-like operator, $\uptau$, must modify the conformal dimension as $\Updelta\mapsto \Updelta+\uptau$ \cite{Vega:2008te,BoschiFilho:2012xr,Branz:2010ub}.

The introduction of an anomalous dimension yields the  exploration of other features of mesons. Indeed, mesons happen to be non-degenerate states, after CSB. For distinguishing them, the anomalous dimension, $\Updelta_P$,  identifies the parity of the mesonic state, here to be considered the $\eta$ pseudoscalar family. It promotes, hence, a  CSB mechanism, by adjusting the bulk mass as 
\begin{align} \label{mass-anomalous}	
\!\!\!\!\!\!\mathcal{M}_5^2=\frac{1}{R^2}\left(\Updelta_0\!+\!\Updelta_P\right)\left(\Updelta_0\!+\!\Updelta_P-4\right) \!-\!s\left(s-4\right), 
\end{align}
where $\Updelta_0$ emulates an effective anomalous dimension, as proposed in details in Ref. \cite{Contreras:2018hbi}. 
Employing Eqs. (\ref{123}, \ref{acao}), one can derive the mass spectrum of the $\eta$ mesonic states.  Under parity, $J^{PC}=0^{-+}$, the $\eta$ mesons Regge trajectory is not invariant. Ref. \cite{Contreras:2018hbi} used the Breitenholner--Freeman limit to show that $\Updelta_P=-1$, for the $\eta$ meson family. Besides, the inequality $\mathcal{M}_5^2R^2\geq-4$ generates stable  pseudoscalar mesonic states.

The equation of motion that rules the $\eta$ pseudoscalar meson family is obtained by varying the action (\ref{123}), yielding   
\begin{eqnarray}\label{eqnmotion}
	 z^5\frac{\partial}{\partial z}\left(\frac{e^{-k^2z^2}}{z} \frac{\partial P}{\partial z}\right) + {e^{-k^2z^2}}\left(3P - z^2\Box P \right)=0.
\end{eqnarray}
One implements the Fourier transform with respect to the $q^\mu$-momentum space, 
\beq\label{fou}
P(q,z)=\frac{1}{(2\pi)^2}\iiiint_{\scalebox{0.6}{Boundary}}\,d^4q\,e^{ix^\mu q_\mu}P(x^\mu,z), 
\eeq
denoting $
P(q,z)=\tilde{P}(q) p(q,z),$ 
where $\tilde{P}(q)$ stands for the sources and $p(q,z)$ is  
the pseudoscalar bulk-to-boundary propagator, with condition $P(q,z_0)=1$.  Therefore, Eq. (\ref{eqnmotion}) yields 
\begin{eqnarray}\label{eqnmotion1}
	z^2 \frac{\partial^2p}{\partial z^2} - (3+2k^2z^2)p \frac{\partial p}{\partial z}+	\left(z^2q^2+4\right)p=0.
\end{eqnarray}
Following well established holographic procedures for mesons \cite{Karch:2006pv,Branz:2010ub,Vega:2009zb}, the mass spectrum can be derived. 
 Eq. \eqref{eqnmotion1} is a Kummer differential equation, whose solutions are expressed by the Kummer confluent hypergeometric function, $F(a,b,x)$, as
\beq\label{etaeq}
    p\left(q,z\right)=\frac{z^2 F\left(1-\mathfrak{q},1,k^2\,z^2\right)}{z_0^2 F\left(1-\mathfrak{q},1,k^2\,z_0^2\right)},
\eeq
where $\mathfrak{q}\equiv \frac{q^2}{4k^2}.$

The on-shell boundary action in the pseudoscalar case is obtained by evaluating the solution (\ref{etaeq}) into Eq.  (\ref{acao}): 
\begin{multline}
\!\!\!\!\!\!\!  I_{\scalebox{0.6}{Boundary}}^\eta = \frac{R^3}{g^2_S}\!\iiiint_{\scalebox{0.6}{Boundary}}\frac{d^4q}{\left(2\pi\right)^4}\left[\frac{e^{-k^2\,z^2}}{z^3}\tilde{P}\left(q\right)\tilde{P}\left(-q\right)\right.\\
  \left.p\left(q,z\right)\frac{\partial p\left(z,-q\right)}{\partial z}\Big\vert_{z=z_0}\right].  
\end{multline}
Therefore the correlator, $\Pi_\eta(q^2)=\frac{\delta^2 I_{\scalebox{0.6}{Boundary}}^\eta}{\delta\tilde{P}\left(-q\right)\delta\tilde{P}\left(q\right)}$, reads \cite{Braga:2015jca} 
\begin{equation}\label{abo}
\Pi_\eta\left(q^2\right)= \frac{R^3e^{-k^2z^2}}{g^2_Sz^3}\frac{\partial p\left(q,z\right)}{\partial z}\Big\vert_{z=z_0}.    
\end{equation}
 Replacing Eq. (\ref{etaeq}) into (\ref{abo}) yields
\begin{multline}
\!\!\!\!\!\!\Pi_\eta\left(q^2\right)\!=\!-\left[\frac{2}{z_0}\!+\!\left(1-\mathfrak{q}\right)\frac{2k^2z_0F\left(2-\mathfrak{q},2,k^2z^2_0\right)}{F\left(\mathfrak{q},1,k^2z^2_0\right)}\right]\\
\times\frac{R^3e^{-k^2z^2_0}}{g_S^2z_0^3}\label{pieta}.     
\end{multline}    
As one regards the on-shell mass condition, $q^2=M_n^2$, the $\eta$ mesons family mass spectrum can be read of the poles of (\ref{pieta}), corresponding to the roots $\xi_n\equiv M_n^2/4k^2$ of the Kummer function in the denominator of Eq. (\ref{pieta}), namely \cite{Cortes:2017lgz}, $
    F\left(1-\xi_n,1,k^2z^2_0\right)=0.$ 
Hence the $\eta$ mesons mass spectrum reads
\begin{equation}
M_n^2=4k^2\xi_n\left(k,z_0,\Updelta_P\right). 	
\label{spector}
\end{equation}
The numerical results are presented in the fourth column of Table \ref{scalarmasses}, when taking  $ k= 0.450$ GeV and $z_0 = 5.0$ GeV${}^{-1}$. 
\begin{figure}[H]
	\centering
	\includegraphics[width=8cm]{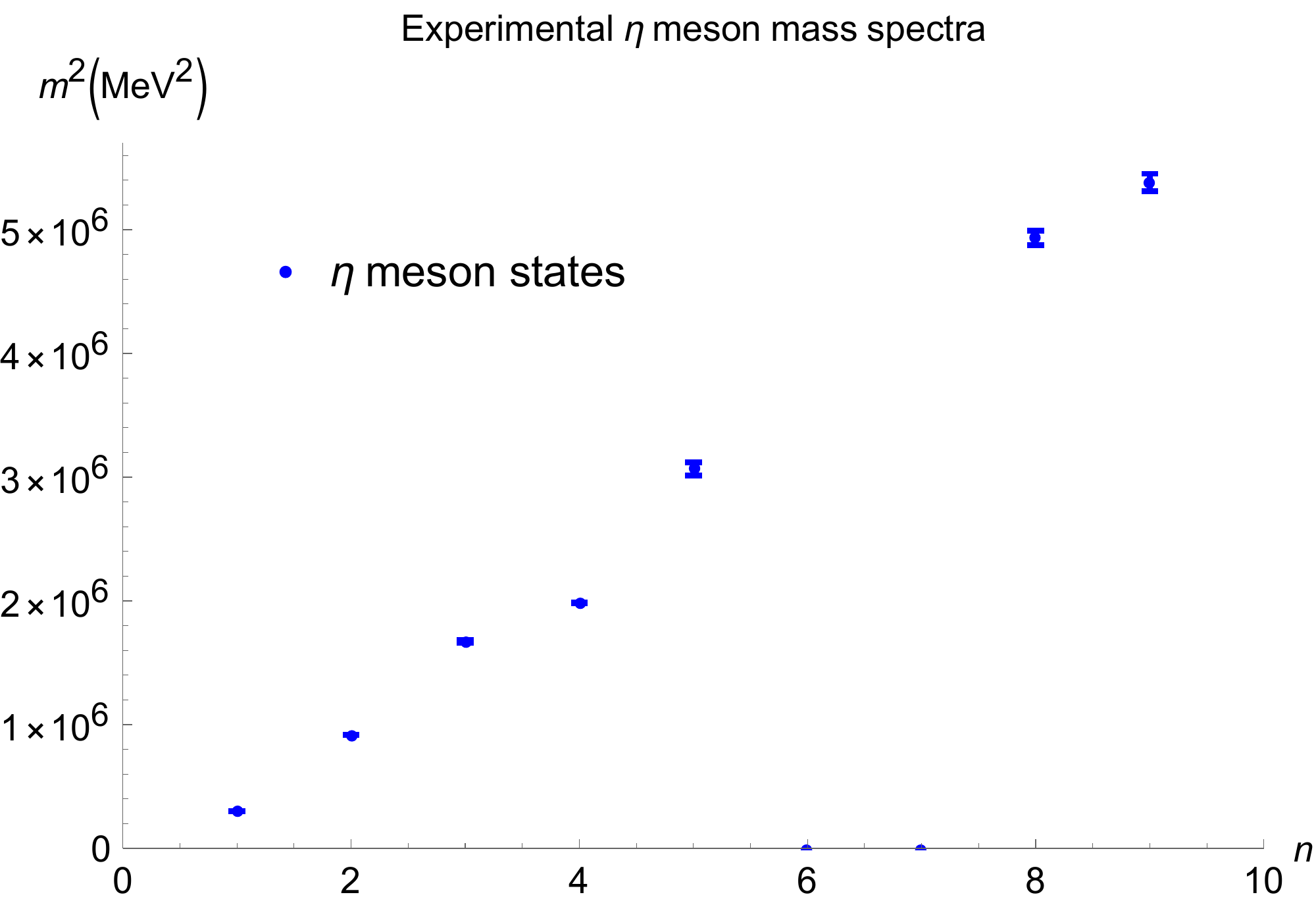}
	\caption{\blt{Experimental squared mass spectrum of the $\eta$ pseudoscalar meson family, for $n=1,\ldots,9$, with error bars, with states $\eta(550)$, $\eta'(958)$,
	 $\eta(1295)$, $\eta(1405)\sim\eta(1475)$, $\eta(1760)$, $\eta(2225)$, $\eta(2320)$ in PDG \cite{pdg}.}}
	\label{figm2xj}
\end{figure}

\blt{\begin{table}[h]
\begin{center}----------- \;$\eta$ pseudoscalar mesons mass spectrum \;------------\medbreak
\begin{tabular}{||c|c||c|c||}
\hline\hline
$n$ & State & $M_{\scalebox{.67}{\textsc{Experimental}}}$ (MeV)  & $M_{\scalebox{.67}{\textsc{Theory}}}$ (MeV) \\
       \hline\hline
\hline
1 &\;$\eta(550)\;$ & $547.862\pm 0.017$ & 975.2   \\ \hline
2 &\;$\eta'(958)\;$ & $957.78 \pm 0.06$ & 1011.1   \\ \hline
3& \;$\eta(1295)\;$& $1294\pm 4$       & 1234.0     \\\hline
4& \;$\eta(1405)/\eta(1475)\;$& $1408.8\pm 1.8 $  & 1455.4    \\\hline
5& \;$\eta(1760)\;$& $1751\pm 15$     & 1829.1   \\\hline
6& \;$\eta(\Box)\;$&    & 1992.6  \\\hline
7& \;$\eta(\Box)\;$&      &  2087.3  \\\hline
8& \;$\eta(2225)\;$& $2221^{+13}_{-10}$      & 2193.6   \\\hline
9& \;$\eta(2320)\;$& $2320\pm 15$      & 2289.4  \\\hline
\hline\hline
\end{tabular}
\caption{\blt{Both the experimental and the AdS/QCD-predicted mass spectrum of the $\eta$ pseudoscalar meson family. The identification of the $\eta$ particle states to their respective $n$ excitation numbers follow the seminal work \cite{Rinaldi:2021dxh}.} } \label{scalarmasses}
\end{center}
\end{table}}\\
\blt{As discussed in Sec. 63 of PDG 2020 \cite{pdg}, the $\eta(1405)$ and the $\eta(1475)$ states might be the same particle \cite{amsler,masoni} and here this identification is implemented. The $\eta(\Box)$ state with $n=6$, in Table \ref{scalarmasses},  might be identified to the $\eta(2010)$, with experimental mass $2010^{+35}_{-60}$ MeV, whereas the $\eta(\Box)$ state with $n=7$ might be identified to the $\eta(2100)$, with experimental mass $2100^{+30+75}_{-24-26}$ MeV \cite{pdg}. The 
$\eta(2100)$ is argued to be alternatively interpreted as a pseudoscalar glueball  or just a Regge excitation of $\eta'$. For more details see, e. g., Refs. \cite{He:2009sb,Godfrey:1998pd,Amsler:2004ps,Klempt:2007cp,Bugg:1999jc}.}

\section{DCE Regge-like trajectories of $\eta$ mesons}\label{ce1}
\label{sec2}
The DCE evaluates correlations amongst the fluctuations of the energy configurations into the physical system to be studied. To portray the system, the energy density -- the time component of the stress-energy-momentum tensor $T_{00}({\bf r})$ -- is the essential ingredient, where ${\bf r}\in\mathbb{R}^m$, for any finite dimension $m$. The correlator 
\beq
\Pi({\bf r})=\int_{\mathbb{R}^m} \,d^m x\, T_{00}(\mathring{\bf r}+{\bf r})T_{00}(\mathring{\bf r})\eeq
 defines the probability distribution  that determines the DCE as  the Shannon's (information) entropy of correlations \cite{Braga:2018fyc}.  
For constructing the DCE, one first calculates the Fourier transform 
\beq\label{fou}
T_{00}({\bf k}) = \frac{1}{(2\pi)^{m/2}}\int_{\mathbb{R}^m}\,d^m x\,T_{00}({\bf r})e^{-i{\bf k}\cdot {\bf r}}.\eeq 
A detector identifies a wave mode, within a volume $d^m{k}$ centered
at ${\bf k}$, with a probability that is proportional to the power spectrum in that
mode, $p({\bf k}\vert d^m{k})\sim {\left|T_{00}({\bf k})\right|^{2}}d^m{k}$ \cite{Gleiser:2018kbq}.  Eq. (\ref{fou}) can be also seen as a probability distribution in momentum space for different wavelengths that impart the generation of correlations across the system. This probability distribution engenders the  modal fraction     
\cite{Gleiser:2012tu}, 
\begin{eqnarray}
\mathcal{T}_{00}({\bf k}) = \frac{\left|T_{00}({\bf k})\right|^{2}}{ \int_{\mathbb{R}^m} \,d^m{k}\, \left|T_{00}({\bf k})\right|^{2}}.\label{modalf}
\end{eqnarray} 
The amount of information to describe $T_{00}$, with respect to wave modes, is computed by the DCE, 
\begin{eqnarray}
{\rm DCE}_{T_{00}}= - \int_{\mathbb{R}^m}\blt{{\mathcal{T}_{00}^{\scalebox{.93}{$\star$}}}({\bf k})\ln  {\mathcal{T}_{00}^{\scalebox{.93}{$\star$}}}}({\bf k})\, d^mk\,,
\label{confige}
\end{eqnarray}
where \blt{$\mathcal{T}_{{00}}^{\scalebox{.93}{$\star$}}({\bf k})=\mathcal{T}_{00}({\bf k})/\mathcal{T}_{{00}}^{\scalebox{.58}{max}}({\bf k})$}, and $\blt{\mathcal{T}_{{00}}^{\scalebox{.58}{max}}({\bf k})}$ is the supremum of \blt{$\mathcal{T}_{00}$} in $\mathbb{R}^m$ wherein the power spectrum  peaks. The DCE has units of nat/unit volume\footnote{Nat denotes the natural unit of information. One nat equals $1/\ln 2$ bits, representing the information that underlies a uniform distribution defined on the real range $[0,e]$.}.  The DCE encodes a scale information, as the power spectrum is represented the Fourier
transform of the 2-point correlator \cite{Gleiser:2018kbq}.

As we are here scrutinizing the $\eta$ pseudoscalar meson family, the value $m=1$ is taken into account in Eqs. (\ref{fou} -- \ref{confige}), representing  the  scale energy, $z$, along the AdS bulk. 
Replacing the Lagrangian (\ref{acao}) into the time component of the  energy-momentum tensor,  
 \begin{equation}
\blt{ \!\!\!\!\!\!\!\!T_{00}\!=\!  \frac{2}{\sqrt{ -g }}\!\! \left[\frac{\partial (\sqrt{-g}{L})}{\partial{g^{00}}} \!-\!\frac{\partial}{\partial{ x^\gamma }}  \frac{\partial (\sqrt{-g} {L})}{\partial\left(\frac{\partial g^{00}}{\partial x^\gamma}\right)}
  \right]},
  \label{em1}
 \end{equation} 
 the expectation value of the energy density, to be used in Eqs. (\ref{fou} -- \ref{confige}), reads \beq\label{t00}
\left\langle T_{00}(z)\right\rangle =\frac{1}{z^4}\,\left[P^{\prime2}(z) + \mathcal{M}_5^2P^2(z)\right]\,.
\eeq
With the energy density (\ref{t00}) of the $\eta$ pseudoscalar meson family in hands, an alternative procedure for deriving their mass spectrum is put into action, solely using the DCE. It is worth to emphasize that this method, that hybridizes AdS/QCD, can be realized as a more accurate one, when compared to pure  AdS/QCD predictions. In fact, it is based on the interpolation of the $\eta$ meson family experimental mass spectrum in PDG \cite{pdg}.

The DCE, using the protocol (\ref{fou} -- \ref{confige}), for the $\eta$ pseudoscalar  meson family, can be numerically computed and the results are shown in Table \ref{cen}. 
\begin{table}[h]
\begin{center}
\begin{tabular}{||c|c|c||}
\hline\hline
       \,$n$ \,&\, State \,&\,CE  \\ \hline\hline
1\,&\,\,$\eta(550)$ \,&\,3.994 \,\\\hline
2\, &\;$\eta'(958)\;$ &\,6.212  \,\\\hline
3\,&\, $\eta(1295)$\,&8.248\,\\\hline
4\,&\, $\eta(1405)\sim\eta(1475)$\,&9.943\,\\\hline
5\,&\, $\eta(1760)$\,&\, 14.540  \,\\\hline
6\,&\, $\eta(\Box)$\,&  23.834\,\\\hline
7\,&\, $\eta(\Box)$\,&\, 28.821 \,\\\hline
8\,&\, $\eta(2225)$\,&\, 35.561 \,\\\hline
9\,&\, $\eta(2320)$\,&\, 42.035 \,\\\hline
10*\,&\, $\eta_{10}$\,&\,  48.474\,\\\hline
11*\,&\, $\eta_{11}$\,&\,  54.007\,\\\hline
12*\,&\, $\eta_{12}$\,&\, 58.464\,\\\hline
\hline\hline
\end{tabular}
\caption{The DCE of the $\eta$ pseudoscalar  meson family. States with an asterisk  represent the members of the $\eta$ meson family whose mass is obtained by the DCE Regge-like trajectories (\ref{itp1}, \ref{itq1}), \blt{respectively for the excitation numbers $n=10, 11, 12$.}}
\label{cen}
\end{center}
\end{table}

\begin{figure}[H]
	\centering
	\includegraphics[width=8.5cm]{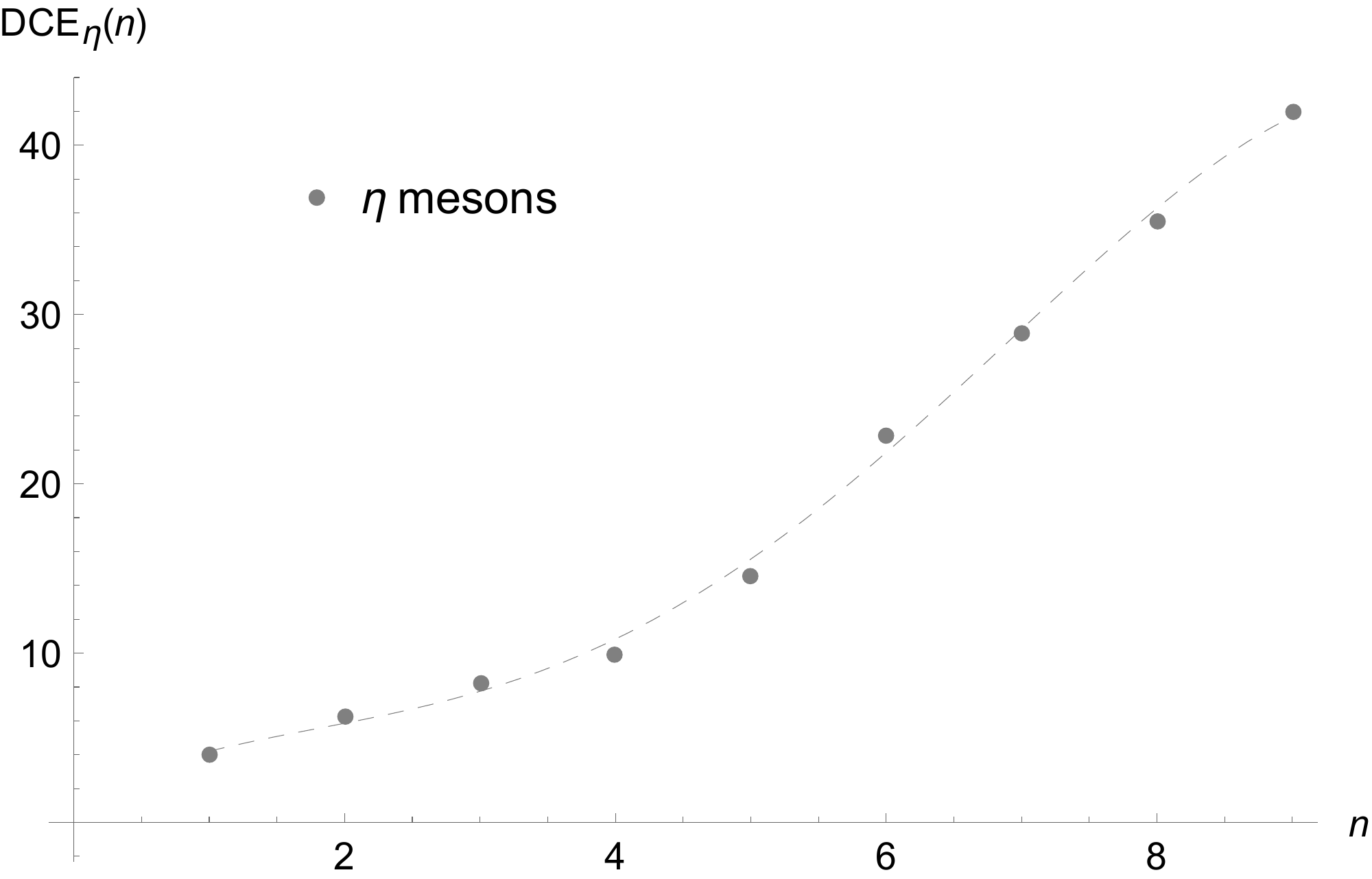}
	\caption{\blt{DCE of the $\eta$ pseudoscalar  meson family, for $n=1,\ldots,9$ (corresponding to the $\eta(550)$, $\eta'(958)$,
	 $\eta(1295)$, $\eta(1405)\sim\eta(1475)$, $\eta(1760)$, $\eta(2225)$, $\eta(2320)$ in PDG \cite{pdg} and to the two states $\eta(\Box)$) as a function of $n$. 
The first form  of DCE Regge-like trajectory is plotted as the gray dashed line.}}
	\label{cen1}
\end{figure}

Standard Regge trajectories AdS/QCD show the light-flavor meson spin to be proportional to the square of the mass spectrum. One can then emulate them, using the intrinsic DCE of the $\eta$ pseudoscalar meson family. For it, the DCE of the $\eta$ pseudoscalar mesonic states can be regarded as a function of the $\eta$ mesons mass spectrum that have been already detected in experiments \cite{pdg}. The interpolation curve of this data consists of the second form of DCE Regge-like trajectories, whereas the first form takes into account the DCE as a function of the $n$ excitation modes of $\eta$ meson states. Fig. \ref{cen1} shows the corresponding results, wherein numerical interpolation  yields the first form of DCE Regge-like trajectory. 
The explicit expression is numerically obtained by interpolation of data in Table \ref{cen}:
\begin{eqnarray}\label{itp1}
 \blt{\!\!\!\!{\rm DCE}_\eta(n)} \!&=&\! \blt{-0.021n^4\!+\!0.377 n^3\!-\!1.604 n^2}\!\nonumber\\&&\blt{+4.155 n+1.297}, 
   \end{eqnarray} \blt{within $\sim 1.5\%$ standard deviation}. 
\begin{figure}[H]
	\centering
	\includegraphics[width=8.5cm]{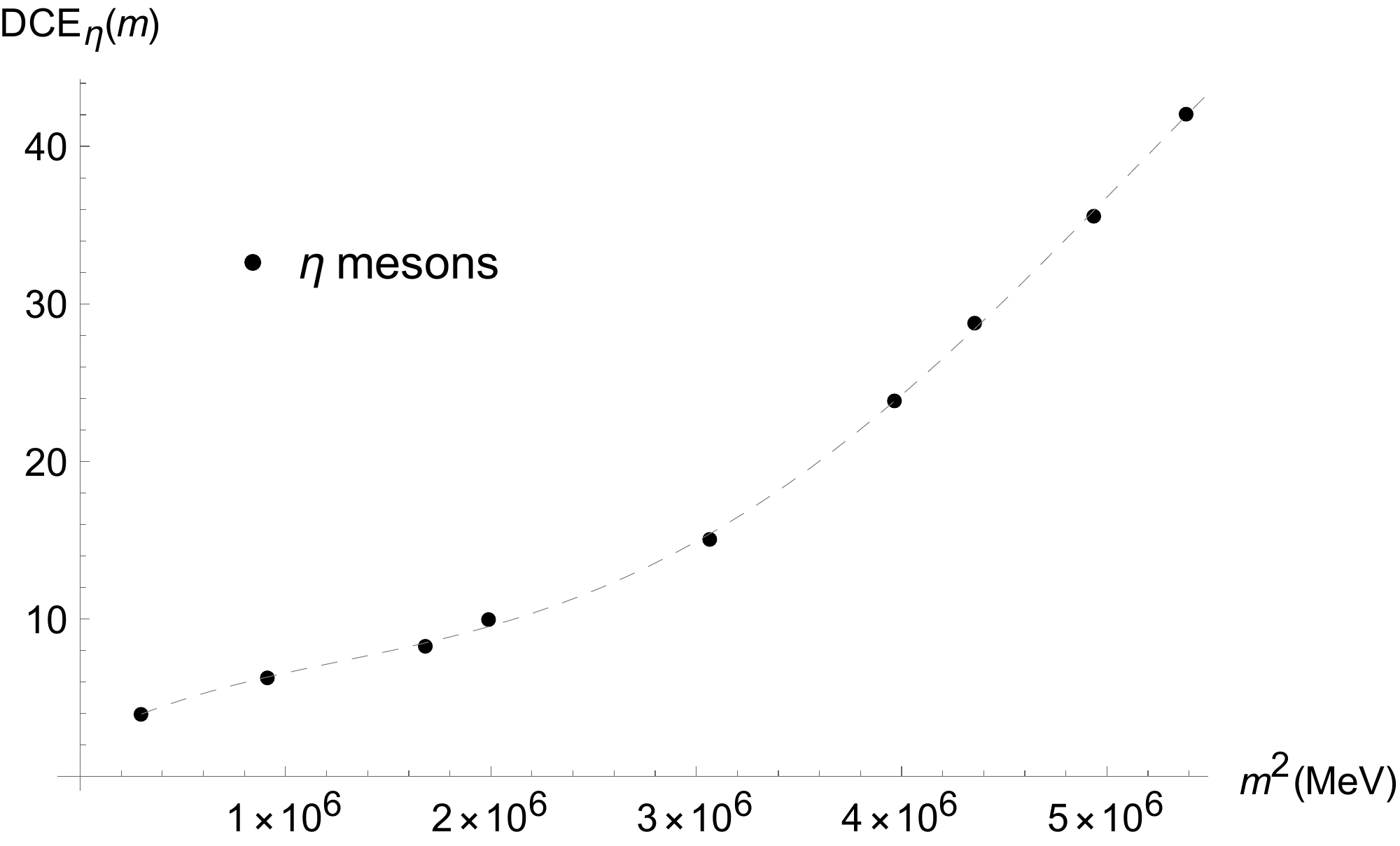}
	\caption{\blt{DCE of the $\eta$ pseudoscalar  meson family, for  $n=1,\ldots,9$ (corresponding to the $\eta(550)$, $\eta'(958)$,
	 $\eta(1295)$, $\eta(1405)\sim\eta(1475)$, $\eta(1760)$, $\eta(2225)$, $\eta(2320)$ in PDG \cite{pdg} and to the two states $\eta(\Box)$) as a function of their (squared) mass. 
The second form  of DCE Regge-like trajectory is plotted as  the gray dashed line.}}
	\label{cem1}
\end{figure}

\blt{As there is no known experimental data associated with the $\eta(\Box)$ states, the hybrid interpolation in Fig. \ref{cem1} takes the experimental masses of $\eta(550)$, $\eta'(958)$,
	 $\eta(1295)$, $\eta(1405)\sim\eta(1475)$, $\eta(1760)$, $\eta(2225)$, $\eta(2320)$ in PDG \cite{pdg} and the AdS/QCD-predicted masses of the two states $\eta(\Box)$, as in Table \ref{scalarmasses}. As already pointed out, one might try to identify the $\eta(\Box)$ state with $n=6$  to $\eta(2010)$, with mass $2010^{+35}_{-60}$ MeV, and the $\eta(\Box)$, $n=7$, state to $\eta(2100)$, whose mass equals $2100^{+30+75}_{-24-26}$ MeV \cite{pdg}. However, as the $\eta(2100)$ may be a pseudoscalar glueball  or a Regge excitation of $\eta'$, taking the $\eta(2100)$ mass in PDG for deriving the DCE Regge-like trajectories is too speculative. Anyway, either using the hybrid interpolation  or the interpolation with the experimental masses of $\eta(2010)$ and $\eta(2100)$  yields a tiny difference of $0.1\%$ in the derived masses of $\eta_{10}, \eta_{11}, \eta_{12}$. Hence, as both methods provide similar masses of the $\eta_{10}, \eta_{11}, \eta_{12}$ states up to $0.1\%$, only the hybrid interpolation in Fig. \ref{cem1} will be taken into account in what follows.}
\noindent The DCE Regge-like trajectory, as a function of the $\eta$ mesons mass, $m$ (MeV), is listed as follows:
\begin{eqnarray}\label{itq1}
\!\!\!\!\!\!\!\!\!\!\!\!\blt{{\rm DCE}_\eta(m)} &=&  \blt{1.476\times \!10^{-25} m^8
 +1.808\times \!10^{-18} m^6} \nonumber\\&&\blt{+ 7.798\times \!10^{-12} m^4 
 - 8.320\times\! 10^{-6} m^2}\nonumber\\&&\qquad\blt{+  5.823},
   \end{eqnarray} \blt{within $\sim0.17\%$ standard deviation}. 
The DCE Regge-like trajectories in Eqs. (\ref{itp1}, \ref{itq1}) illustrate several important features about the $\eta$ pseudoscalar  meson family. Using  (\ref{itp1}, \ref{itq1}) makes one to compute the DCE of members of the $\eta$ pseudoscalar meson family, for  $\blt{n\geq 10}$. They consist of the first next  
states of the    $\eta$ pseudoscalar meson family to be detected in experiments. 

\blt{As illustrated in Table \ref{cen}, replacing  $n=10$ in Eq. (\ref{itp1}) yields the DCE equal to 48.474. Then, substituting this value in the left-hand side of Eq. (\ref{itq1}), and solving for $m$, one obtains for the tenth member, $\eta_{10}$, of the $\eta$ pseudoscalar meson family, the mass $m_{\eta_{10}}= 2428.6$ MeV. Implementing a similar procedure for $n=11$, one derives the mass of the eleventh member, $\eta_{11}$, as $m_{\eta_{11}}= 2533.1$ MeV. Looking at PDG \cite{pdg}, one can attempt to identify the $X(2632)$ meson, having mass $m=2635.2\pm 3.3 \text{MeV}$, to the $\eta_{11}$. 
Besides, replacing  $n=12$ in Eq. (\ref{itp1}) yields the DCE equals to 58.464. Therefore, putting back this value into the left-hand side of Eq. (\ref{itq1}), solving for $m$ yields the mass  \blt{$m_{\eta_{12}}= 2695.9$ MeV}, for the  $\eta_{12}$ member of the $\eta$ pseudoscalar meson family. One can speculate that the $\eta_{12}$ mesonic state corresponds to the already detected $X(2680)$, with mass $2676\pm 27$ MeV \cite{pdg}. }

\section{Concluding remarks}\label{iv}

We regarded a useful technique, hybridizing DCE to AdS/QCD, to derive the mass spectrum of $\eta$ pseudoscalar meson states. with higher excitation number. For it, the quadratic dilaton model with anomalous dimension was employed. The compactness of DCE methods, used in Sec. \ref{sec2}, is here advantageous for at least two main reasons. The first one is the avoidance of computational issues that are common in approaching AdS/QCD phenomenology. The second one resides in hybridizing DCE to AdS/QCD. In this case, obtaining the mass spectrum of the next generation of $\eta$ mesons to be detected takes into account the \blt{nine} first $\eta$ mesons masses that were experimentally detected \cite{pdg}, as illustrated in Fig. \ref{cem1} and analytically approximated by Eq. (\ref{itq1}) with a very good accuracy. With the aid of data in Fig. \ref{cen1} and Table \ref{cen}, another form  of DCE Regge-like trajectory (\ref{itp1}) completes the necessary ingredients for deriving the mass spectrum of $\eta$ pseudoscalar meson states. 
The derived mass spectrum agrees  with 
 meson states candidates, recently listed in PDG 2020 \cite{pdg}, as already comprehensively discussed in the last paragraph of Sec. \ref{sec2}. 

According to Table \ref{cen} and Fig. \ref{cen1}, $\eta$ pseudoscalar meson states with lower $n$ excitation number have higher configurational stability, from the DCE point of view. Lower values of the DCE can point to more predominant $\eta$ meson  states. Similarly, the more massive the $\eta$ pseudoscalar meson state, the higher its configurational instability is. Hence, more massive $\eta$ pseudoscalar meson states  are more unstable from  the point of view of the DCE, being also less predominant and prevalent $\eta$ meson states. This interpretation also corroborates with the number of events, that characterizes each $\eta$ meson, already detected by experiments \cite{pdg}. Besides the peculiarities of producing each $\eta$ meson as a by-product of precise interactions and reactions, the DCE might also suggest  the phenomenological prevalence of states in the $\eta$ pseudoscalar meson family.

\medbreak
\paragraph*{Acknowledgments:}RdR is grateful to FAPESP (Grant No. 2017/18897-8) and to the National Council for Scientific and Technological Development -- CNPq (Grants No. 303390/2019-0 and No. 406134/2018-9), for partial financial support. 

\end{document}